\documentclass[a4paper,11pt]{article}

\usepackage{pos}
\usepackage{lipsum} %package to generate placeholder text in the following
\usepackage{gensymb}

\newlength{\bibitemsep}
\newlength{\bibparskip}
\let\oldthebibliography\thebibliography
\renewcommand\thebibliography[1]{%
  \oldthebibliography{#1}%
  \setlength{\parskip}{\bibitemsep}%
  \setlength{\itemsep}{\bibparskip}%
}

\title{Sensitivity of the IceCube-Gen2 Surface Array for Cosmic-Ray Anisotropy Studies}

\ShortTitle{Sensitivity of the IceCube-Gen2 surface array for CR anisotropy}

\author{The IceCube-Gen2 Collaboration \\{\normalsize \normalfont(a complete list of authors can be found at the end of the proceedings)}\\}

\emailAdd{wenjie.hou@kit.edu}

\abstract{

The energy of the transition from Galactic to extra-galactic origin of cosmic rays is one of the major unresolved issues of cosmic-ray physics. However, strong constraints can be obtained from studying the anisotropy in the arrival directions of cosmic rays. The sensitivity to cosmic-ray anisotropy is, in particular, a matter of statistics. Recently, the cosmic ray anisotropy measurements in the TeV to PeV energy range were updated from IceCube using 11 years of data. The IceCube-Gen2 surface array will cover an area about 8 times larger than the existing IceTop surface array with a corresponding increase in statistics and capability to investigate cosmic-ray anisotropy with higher sensitivity. In this contribution, we present details on the performed simulation studies and sensitivity to the cosmic-ray anisotropy signal for the IceCube-Gen2 surface array.
\vspace{4mm}
{\bfseries Corresponding authors:}
Wenjie Hou$^{1*}$\\
{$^{1}$ \itshape Karlsruhe Institute of Technology (KIT), Institute for Astroparticle Physics, D-76021 Karlsruhe, Germany}\\[4mm]
$^*$ Presenter

\ConferenceLogo{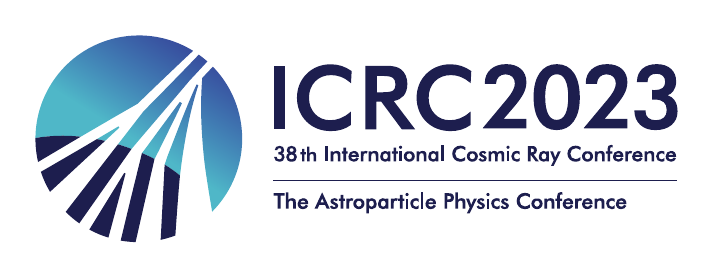}

\FullConference{The 38th International Cosmic Ray Conference (ICRC2023)\\ 26 July -- 3 August, 2023\\ Nagoya, Japan}
}

\begin{document}

\maketitle

%%%%%%%%%%%%%%%%%%%%%%%%%%%%%%%%%%%%%%%% Section 1 %%%%%%%%%%%%%%%%%%%%%%%%%%%%%%%%%%%%%%%%%%
\section{Introduction}\label{sec1}
One of the most significant unresolved questions in cosmic ray (CR) physics pertains to the energy at which the transition from Galactic to extragalactic cosmic rays occurs. This transition holds the key to understanding some of the issues regarding the origin of CRs, as it is believed to mark the point where the sources of CRs shift from the Milky Way to extragalactic systems. Pinpointing the exact energy of this transition remains a challenge, as the trajectories of CRs are signiﬁcantly influenced by the magnetic fields present in the Galaxy, making it difficult to trace individual CRs back to their specific origins.

However, constraints can be obtained by studying the large-scale anisotropy in the CR arrival directions. In the past few decades, several experiments have provided long-term and significant observations of a subtle sidereal anisotropy across a wide energy range from 1 TeV to 100 EeV. These ground-based experiments, located in both the northern and southern hemispheres, have detected large-scale anisotropy in CRs with a high level of statistical significance \cite{Abeysekara:2018uu,Chiavassa:2015kit,Alekseenko:2009itae,Aglietta:2009INAF,Ambrosio:2003un,Guillian:2005ko,Abdo:2009nasw,Bartoli:2015nu,Amenomori:2005hk,Aartsen:2013au,Aartsen:2016au,Aab:2020nu,Gao:2021su}. The magnitude of the observed large-scale anisotropy, ranging from $10^{-4}$ to $10^{-2}$, combined with its energy dependence and angular structure, hints at changes in the origin of the CRs versus energy.

However, statistically significant measurements of the projected dipole amplitude are missing in the energy range between 2 PeV and 8 EeV. The Pierre Auger Observatory has not yet observe a significant dipole below 8 EeV \cite{Aab:2020nu}, and KASCADE-Grande did not find evidence for a dipole anisotropy in the data with median energy from 2.7 PeV to 33 PeV \cite{Chiavassa:2015kit,Chiavassa:2019KIT}. Therefore, both collaborations set upper limits at $99\%$ confidence on the reconstructed (projected) dipole. IceCube-Gen2, along with its surface array, will be capable of filling a portion of the energy gap between 1 PeV and approximately 100 PeV.

IceCube-Gen2, the next generation of the IceCube Neutrino Observatory \cite{Aartsen:2021au}, will be located at the South Pole with a mission to detect high-energy and ultra-high energy cosmic rays and neutrinos, study hadronic interactions, and contribute to the broader field of astroparticle physics. It will incorporate three detector arrays: a deep optical array, a radio array, and a hybrid surface array \cite{Alan:2021ud}. The hybrid surface array combines scintillator panels and radio antennas in surface stations. Currently, the IceCube-Gen2 surface array is in the stage of technical design, but CORSIKA simulations of air shower detection have already been performed \cite{Alan:2021ud}.

Here, we present the air shower reconstruction efficiency for the scintillators of the IceCube-Gen2 surface array, describe the Monte Carlo studies of the CR arrival directions and show the sensitivity of the surface array to large-scale anisotropy of the CRs based on simulated events.  

%%%%%%%%%%%%%%%%%%%%%%%%%%%%%%%%%%%%%%%% Section 2 %%%%%%%%%%%%%%%%%%%%%%%%%%%%%%%%%%%%%%%%%%
\section{Reconstruction efficiency of IceCube-Gen2 surface array}\label{sec2}

The planned surface area covered by the IceCube-Gen2 surface array is 6.6 km$^2$ (depends on the definition of the containment cuts). The corresponding CORSIKA simulations of the scintillator array response were performed for proton- and iron- induced air showers with $4\leq\log_{10} (E/\mathrm{GeV})\leq8$ and zenith angles $(\theta)$ up to $51\degree$, and $4\leq\log_{10} (E/\mathrm{GeV})\leq7.5$ with $\theta$ up to $63\degree$ \cite{Alan:2021ud}. To simulate a large sample of contained events, the shower cores were randomly distributed within a 1.5 km radius from the center of the surface array. The detector response was then simulated for the scintillators, where the secondary particles on the ground were injected into the scintillator panels.

To show realistic capabilities of the IceCube-Gen2 surface array, we perform a selection process that involves choosing true air shower core locations (x, y) within a distance of 100 meters from the polygonal edge of the array taking into account the air shower footprints. The scintillator array is fully efficient (100$\%$) in triggering on air showers above 0.5 PeV for vertical events ($\theta=0$). Considering the scintillator triggered multiplicities $\geq5$,
%%%%%% Figure 1 %%%%%%
\begin{figure}[tbp]
\centering
\includegraphics[width=0.5\textwidth]{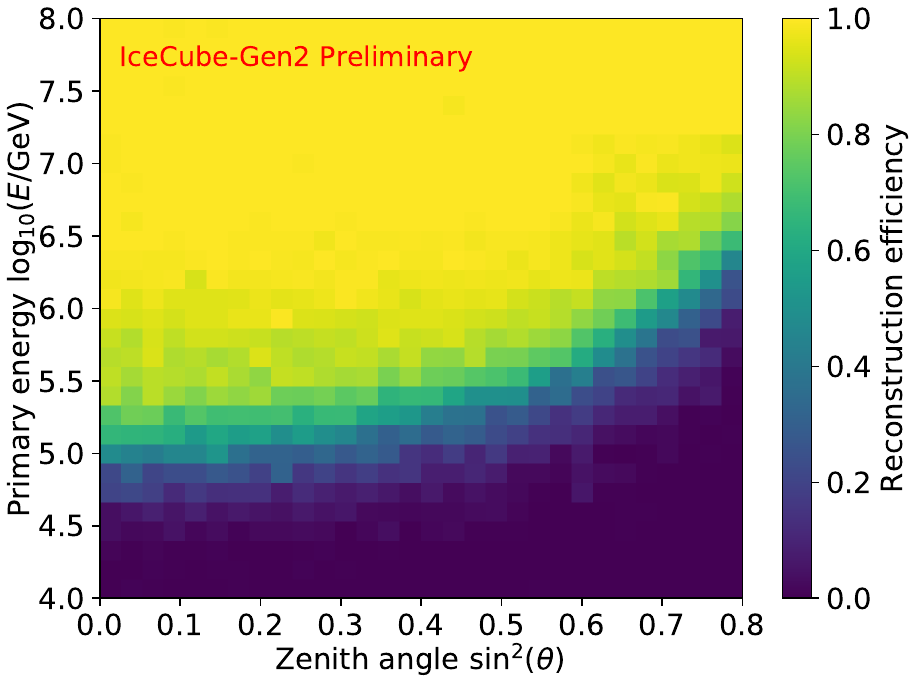}\hfill
\includegraphics[width=0.5\textwidth]{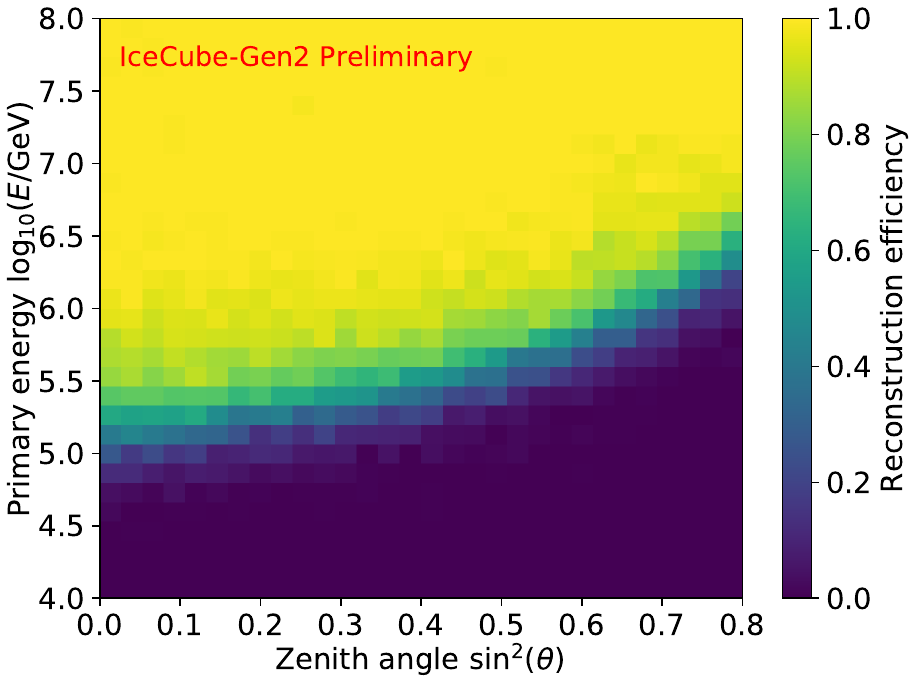}
\caption{2D histograms of the reconstruction efficiency for proton- (left) and iron- (right) induced air shower for scintillators in the IceCube-Gen2 surface array, with energy range from $10^6$ GeV to $10^8$ GeV, zenith angle up to $63\degree$ in $\sin^2{\theta}$ scale, and scintillator multiplicity $\geq5$ to trigger the event.}
\label{fig1}
\end{figure}
we show the reconstruction efficiencies for proton and iron primaries and make the histograms for these two CR primary particles, fill in the missing region $0.6\leq\sin^2(\theta)\leq0.8$ and $7.5\leq\log_{10}(E/\mathrm{GeV})\leq8$ assuming a $100\%$ efficiency  (see Figure \ref{fig1}), and fit the unbroken histograms with a modified error function (2 dimensional) as a function of the energy $E$ and the zenith angle with $z=\sin^2{\theta}$:
%%%%%% Equation 1 %%%%%%
\begin{equation}\label{eq1}
\begin{split}
\epsilon (E, z)=\frac{1}{2}\left[1+\text{erf}\left(\frac{E-(\mu_{0}+\mu_{1}z+\mu_{2}z^{2}+\mu_{3}z^{3})}{\sigma_{0}+\sigma_{1}z}\right)\right].
\end{split}
\end{equation}
The parameters for both proton and iron in Eq.\eqref{eq1} are listed in Table~\ref{tab1}. We estimate the reconstruction efficiencies for helium, nitrogen, and aluminum using the natural logarithm of their mass number, denoted as $\ln{A}$. This estimation is based on the logarithmic mass dependence of the cosmic-ray primaries. The reconstruction efficiency for helium, nitrogen, and aluminum can be written as
%%%%%% Equation 2 %%%%%%
\begin{equation}\label{eq2}
\begin{split}
\epsilon(A_{i},E,z) = \frac{\ln A_{i}}{\ln A_{\rm Fe} - \ln A_{\rm P}}\left[\epsilon_{\rm Fe}(E,z) - \epsilon_{\rm P}(E,z)\right] + \epsilon_{\rm P}(E,z),
\end{split}
\end{equation}
where $i$ ranges from 1 to 3, representing the estimated helium, nitrogen, and aluminum, respectively. 
%%%%%% Table1 %%%%%%
\begin{table}[b]
\centering
\footnotesize
\begin{tabular}{l @{\hspace{10pt}} c @{\hspace{10pt}} c @{\hspace{10pt}} c @{\hspace{10pt}} c @{\hspace{10pt}} c @{\hspace{10pt}} c} 
\hline\hline
Particle & $\mu_{0}$ & $\mu_{1}$ & $\mu_{2}$ & $\mu_{3}$ & $\sigma_{0}$ & $\sigma_{1}$\\
\hline
Proton & $5.017\pm0.008$ & $0.711\pm0.083$ & $-1.791\pm0.240$ & $3.824\pm0.198$ & $0.483\pm0.008$ & $0.001\pm0.000$\\ 
Iron & $5.177\pm0.006$ & $0.753\pm0.063$ & $-0.665\pm0.182$ & $2.073\pm0.149$ & $0.404\pm0.006$ & $-0.067\pm0.012$\\ 
\hline
\end{tabular}
\caption{Parameters in the fit function \eqref{eq1} of cosmic ray reconstruction efficiency of the IceCube-Gen2 surface array. Only proton and iron are covered in this table.
}
\label{tab1}
\end{table}
The contour lines represent $50\%$ and $98\%$ of the reconstruction efficiency using Eq.(\ref{eq1}), as shown in Figure \ref{fig2}. The total efficiency of all particles utilizes Eq.(\ref{eq1}), (\ref{eq2}) and the H4a flux model \cite{Gaisser:2012du}.
%%%%%%%%%%%%%%%%% Figure 2&3 %%%%%%%%%%%%%%%%%
\begin{figure}[tbp]
\centering
\begin{minipage}[c]{0.49\textwidth}
\centering
\includegraphics[width=\linewidth]{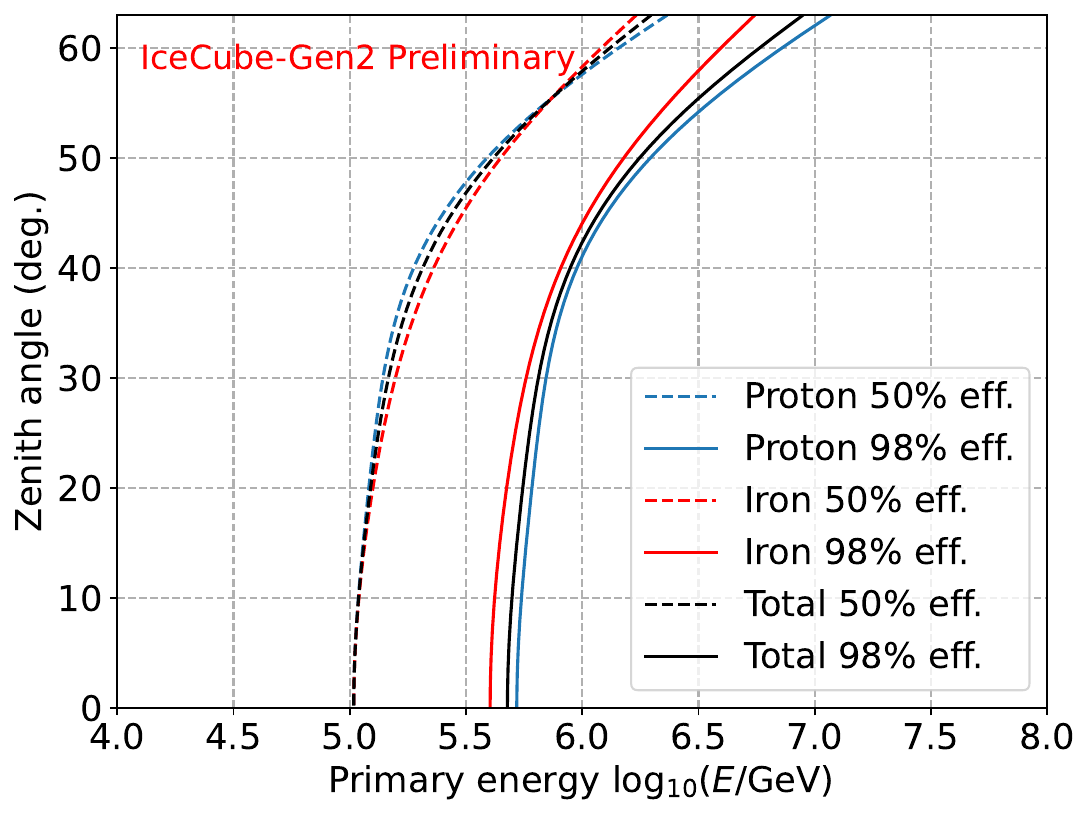}
\caption{Contour lines with $50\%$ and $98\%$ reconstruction efficiency of the IceCube-Gen2 surface array are shown for proton and iron, The total efficiency of all particles utilizes the H4a flux model \cite{Sommers:2001UU}.}
\label{fig2}
\end{minipage}\hfill
\begin{minipage}[c]{0.498\textwidth}
\centering
\includegraphics[width=\linewidth]{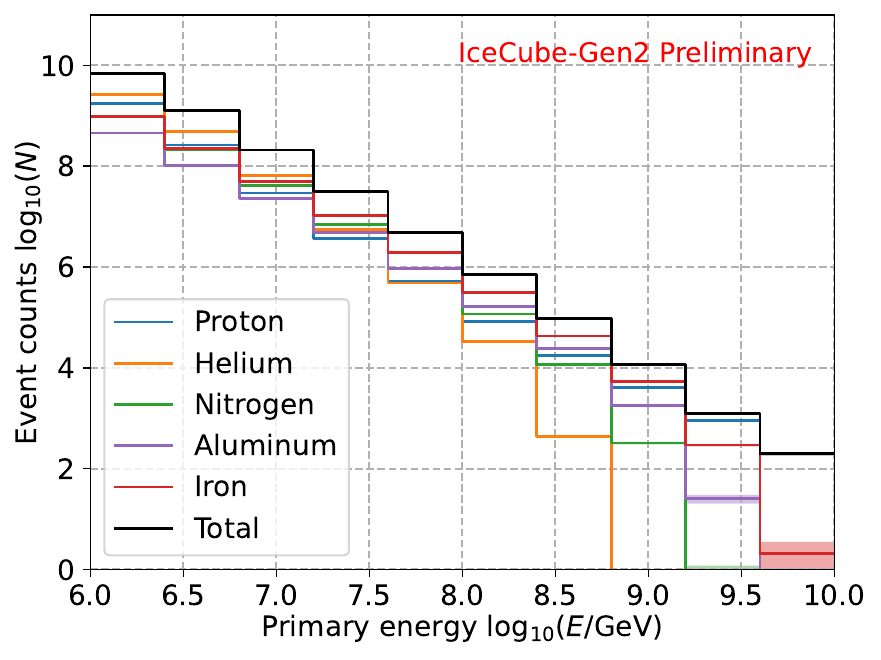}
\caption{Histogram of the event counts of the IceCube-Gen2 surface array with a 10-year exposure, considering the H4a flux model and the reconstruction efficiency with 3 additional energy bins.}
\label{fig3}
\end{minipage}
\end{figure}

The scintillators will be the most sensitive component of the surface hardware of IceCube-Gen2 and thereby determine the energy threshold necessary for cosmic-ray detection. In the resolution analysis of the reconstructed arrival direction \cite{Alan:2021ud} it is observed that, at threshold energies of approximately 1 PeV, the direction can be reconstructed with an accuracy of a few degrees for air showers with zenith angles up to 45$\degree$. Above 10 PeV the angular resolution reaches the sub-degree level. The accurate estimation of the shower geometry and the reliable determination of the arrival direction make it highly valuable for studying cosmic-ray anisotropy.

%%%%%%%%%%%%%%%%%%%%%%%%%%%%%%%%%%%%%%%% Section 3 %%%%%%%%%%%%%%%%%%%%%%%%%%%%%%%%%%%%%%%%%%
\section{Monte-Carlo simulation of cosmic-ray arrival directions}\label{sec3}

We now apply the reconstruction efficiency to simulate the arrival directions for the IceCube-Gen2 surface array with a 10-year exposure. The simulation is divided into 7 energy bins ranging from $10^6$ GeV to $10^{8.8}$ GeV with a bin size of 0.4 in $\log_{10}(E/\mathrm{GeV})$. In each energy bin, we inject 15 different dipoles at declinations ranging from $-80\degree$ to $80\degree$. In total, we have 1785 injected dipoles and their corresponding sky maps. To get a precise and stable value of the reconstructed dipole amplitude we choose dipole amplitudes depending on the declination, focusing on amplitudes that can be detected at $>10\sigma$ within 10 years of operation. Otherwise, it would be required to simulate the same dipole repeatedly to adequately capture the uncertainties.

Using the H4a flux model \cite{Gaisser:2012du}, the number of arrival directions in each energy bin can be calculated by integrating the 2D function in Eq.(\ref{eq1}) and (\ref{eq2}) of reconstruction efficiency and the number of cosmic rays with 10 years of exposure of the IceCube-Gen2 surface array. This calculation is a function of $E$ and $z$. Therefore, we have the event counts for each of the 5 primary components in the energy bins (see Figure \ref{fig3}):
%%%%%% Equation 3 %%%%%%
\begin{equation}\label{eq3}
\begin{split}
N_{\mathrm{eff},ij}=\int^{\theta_{\mathrm{max}}}_{\theta_{\mathrm{min}}}\int^{E_{j}}_{E_{j-1}}\epsilon_{i}\left(E, \theta\right)N_{\mathrm{H4a},i}\left(E\right)dEd\theta,
\end{split}
\end{equation}
where $i$ ranges from 1 to 5 for the 5 primary particles, $j$ ranges from 1 to 8 for the edges of the 7 energy bins, and $N_{\mathrm{H4a}, i}$ represents the event counts for each of the 5 primary particles using the H4a flux model. In total, we have 8.355 billion simulated cosmic ray events over all 7 energy bins. Taking into account the ratio of different primaries weighted by the mass-dependent H4a flux expressions \cite{Gaisser:2012du} and Eq.\eqref{eq1} and \eqref{eq2}, the total efficiency averaged over all primaries is written as 
%%%%%% Equation 4 %%%%%%
\begin{equation}\label{eq4}
\begin{split}
\epsilon_{\mathrm{tot}}\left(E,\theta\right)=\frac{\Sigma^{5}_{i=1}\epsilon_{i}\left(E, \theta\right)\phi_{\mathrm{H4a},i}\left(E\right)}{\Sigma^{5}_{i=1}\phi_{\mathrm{H4a},i}\left(E\right)}.
\end{split}
\end{equation}

Next, we randomly inject the CR arrival directions using relative acceptance of the detector for all energy bins and all zenith angles below threshold and scan over the dipole declination from $-80\degree$ to $80\degree$ with bin size of $10\degree$. The amplitudes are chosen from $7\times10^{-3}$ up to $9.56\times10^{-1}$ for the injected dipole with different energy bins covering different ranges of the amplitude. The distribution function of the CRs is a function of exposure of the IceCube-Gen2 surface array,
%%%%%% Equation 5 %%%%%%
\begin{equation}\label{eq5}
\begin{split}
\omega_{\mathrm{Gen2}}\left(E, \theta, \theta_{\mathrm{max}}, \mathcal{A}, \delta_{\mathrm{d}}, \delta_{\mathrm{Gen2}}\right)=\omega\left(\theta, \theta_{\mathrm{max}}, \delta_{\mathrm{Gen2}}\right)\times \epsilon_{\mathrm{tot}}\left(E, \theta\right)\times \mathcal{D}\left(\mathcal{A}, \delta_{\mathrm{d}}, \delta_{\mathrm{Gen2}}\right),
\end{split}
\end{equation}
where $\delta_{\mathrm{d}}$ denotes the dipole declination, $\omega$ is the relative exposure of an observatory at declination $\delta_{\mathrm{Gen2}}=-89.99\degree$ without detector efficiencies. The function $\mathcal{D}\left(\mathcal{A}, \delta_{\mathrm{d}}, \delta_{\mathrm{Gen2}}\right)$ provides the dipole distribution, taking into account the relative orientation of the IceCube-Gen2 surface array to the dipole.

%%%%%%%%%%%%%%%%%%%%%%%%%%%%%%%%%%%%%%%% Section 4 %%%%%%%%%%%%%%%%%%%%%%%%%%%%%%%%%%%%%%%%%%
\section{Sensitivity to the cosmic ray anisotropy}\label{sec4}

To assess the sensitivity of a partial-sky coverage observatory to a dipole anisotropy, it is essential to compare the actual sky map of injected CR arrival directions (data map with simulated dipole in Eq.(\ref{eq5})) with a sky map that reflects the detector's response to a isotropic CR flux (reference map without dipole). Note that the reference map, with dipole distribution function $\mathcal{D}\left(\mathcal{A}, \delta_{\mathrm{d}}, \delta_{\mathrm{Gen2}}\right)=1$, is not itself isotropic. This comparison allows us to evaluate the deviation from isotropy and quantify the presence of the dipole anisotropy in the simulated data by Eq.\ref{eq5}. Therefore, the residual between the data maps and the reference maps, which are obtained by normalizing each declination band independently, is sensitive only to anisotropy in right ascension ($\alpha$). In particular, the fitting of the dipole's projection in the equatorial plane with the first harmonic allows for a direct measurement of a dipolar cosmic ray distribution.

%%%%%%%%%%%%%%%%%%%%%% Figure 4 %%%%%%%%%%%%%%%%%%%%%%%
\begin{figure}[tbp]
\centering
\includegraphics[width=0.496\textwidth]{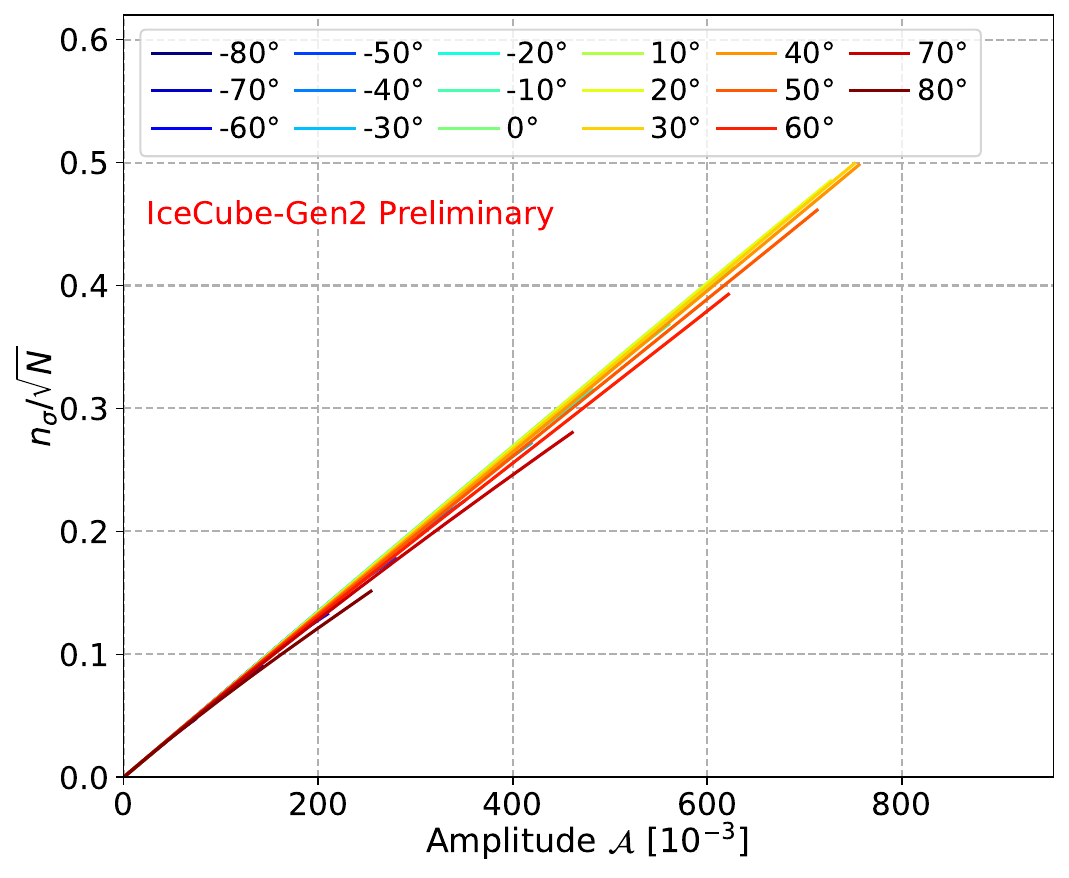}\hfill
\includegraphics[width=0.496\textwidth]{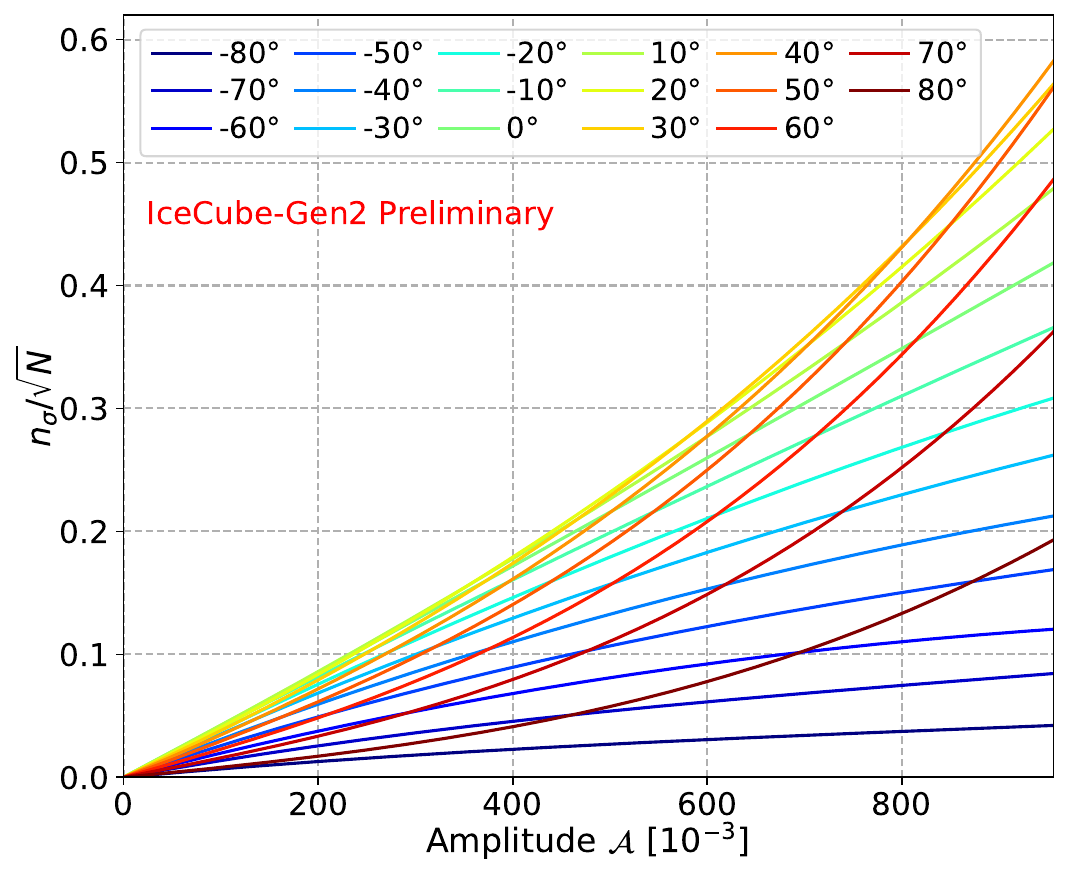}
\caption{The fitting curves of the reconstructed dipole (left) and the corresponding true dipole (right) are shown based on the sampled data points $(n_{\sigma}/\sqrt{N}, \mathcal{A})$ obtained from the simulation in Section \ref{sec3}.}
\label{fig4}
\end{figure}

Based on the Monte-Carlo simulation as described in \textsection\ref{sec3}, we generate sky maps of relative intensity $I=(N_{\mathrm{pix},i}-\left<N\right>_{\mathrm{pix},i})/\left<N\right>_{\mathrm{pix},i}$, where $N_{\mathrm{pix},i}$ represents the data map while $\left<N\right>_{\mathrm{pix},i}$ represents the background map. Then, we perform a one-dimensional (1D) projection of the sky map and fit it with a first harmonic function $\mathcal{A_{\mathrm{reco}}}\cos{(n(\alpha-\phi))}+B$, where $\mathcal{A_{\mathrm{reco}}}$ is the amplitude of the reconstructed dipole, $\phi$ is the phase, and $B$ is a constant. The reconstruction ratio $\mathcal{A_{\mathrm{reco}}}/\mathcal{A}$ in the majority of the cases are below $80\%$ due to the limited field of view (FoV) of the surface array.

\begin{figure}[tbp]
\centering
\begin{minipage}[c]{0.486\textwidth}
\centering
\includegraphics[width=\linewidth]{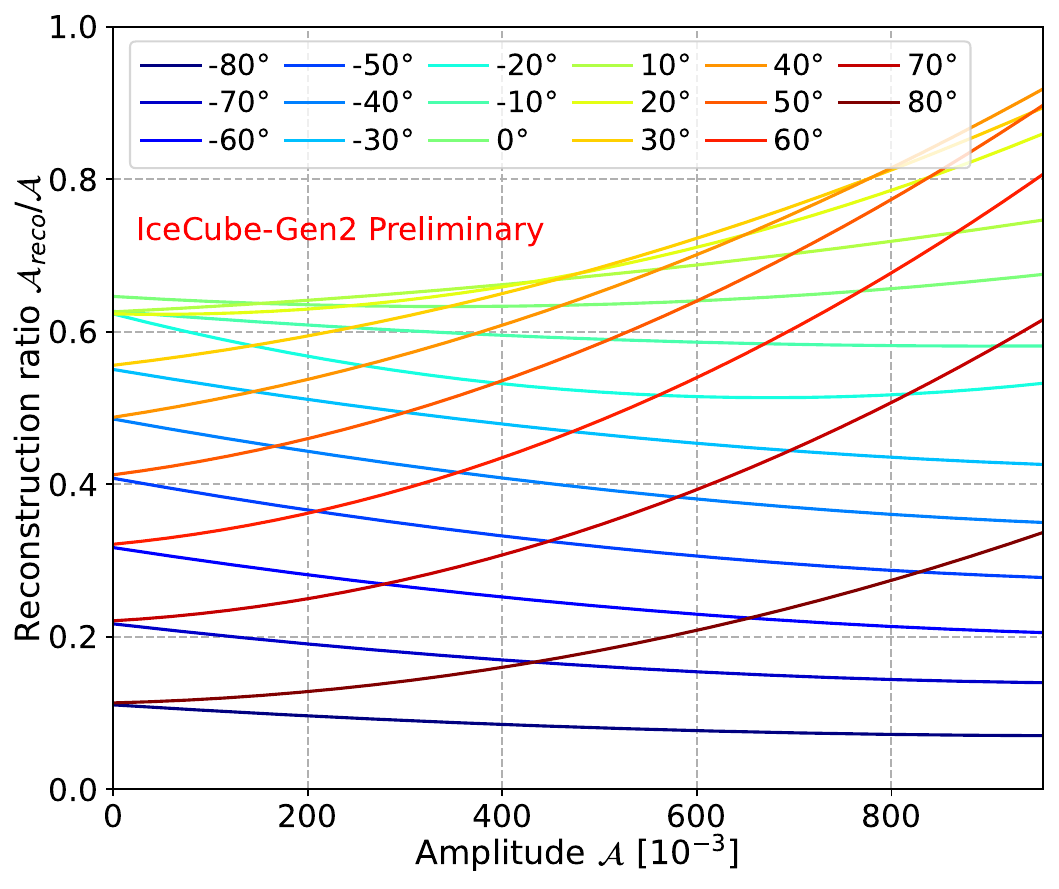}
\caption{Ratio between reconstructed dipole and true dipole in the Monte-Carlo simulation. The amplitudes are chosen from $7\times10^{-3}$ up to $9.56\times10^{-1}$ for the true dipole. 
However, different energy bins cover different ranges of the true dipole amplitude. The curves show the propagation of the ratio with the declination of the true dipole from -80 degree to 80 degree with angle bin size of 10 degree.}
\label{fig5}
\end{minipage}\hfill
\begin{minipage}[c]{0.495\textwidth}
\centering
\includegraphics[width=\linewidth]{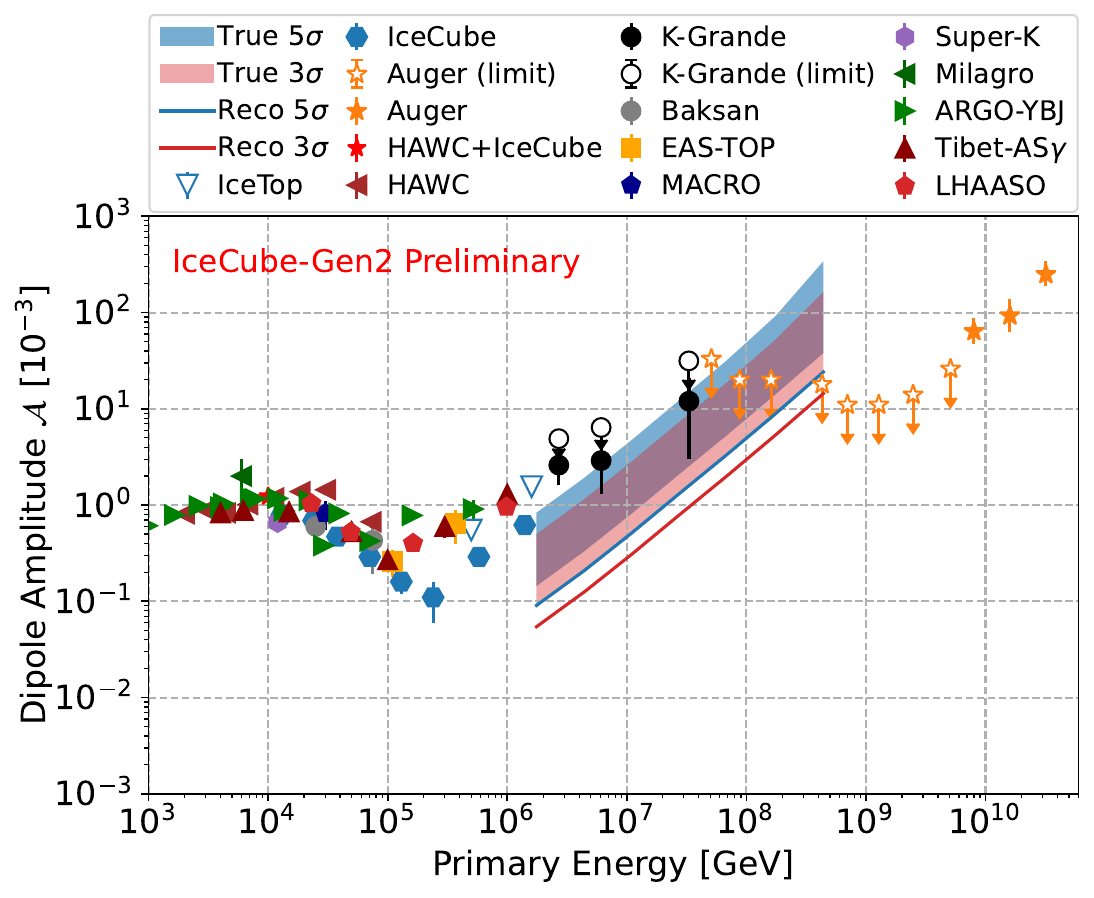}
\caption{The curves show the $3\sigma$ and $5\sigma$ sensitivity of the IceCube-Gen2 surface array to the reconstructed dipole, bands show the corresponding true dipole of the $3\sigma$ and $5\sigma$ reconstructed dipole. The data points shown are reconstructed dipole amplitudes from the various experiments \cite{Abeysekara:2018uu,Chiavassa:2015kit,Alekseenko:2009itae,Aglietta:2009INAF,Ambrosio:2003un,Guillian:2005ko,Abdo:2009nasw,Bartoli:2015nu,Amenomori:2005hk,Aartsen:2013au,Aartsen:2016au,Aab:2020nu,Gao:2021su}. The upper limits of KASCADE-Grande and Auger are from 2.7 PeV to 8 EeV, where the energy gap exists.}
\label{fig6}
\end{minipage}
\label{fig:both}
\end{figure}

To assess the significance of a dipole deviation from isotropy, we consider a null hypothesis in number of sigmas $n_{\sigma}=\left(\mathcal{A_{\mathrm{reco}}}-\mathcal{A_{\mathrm{hypo}}}\right)/\sigma_{\mathcal{A}}$, where $\mathcal{A_{\mathrm{hypo}}}$ is set to 0, representing the expected value under the null hypothesis (different from the full-sky case \cite{Sommers:2001UU}). We assume that the $\sigma_{\mathcal{A}}$ of the dipole amplitude reconstruction is proportional to $1/\sqrt{N}$ verified through both Monte Carlo simulation and the covariance matrix in the first harmonic fit of the 1D projected map. Therefore, the sensitivity function (number of sigmas) can be expressed as follows:
%%%%%% Equation 6 %%%%%%
\begin{equation}\label{eq6}
\begin{split}
n_{\sigma}=\mathcal{S}\left(\mathcal{A}, E, \theta_{\mathrm{max}}, \delta_{\mathrm{d}},\delta_{\mathrm{Gen2}}\right)\mathcal{A}\sqrt{N},
\end{split}
\end{equation}
where $\mathcal{S}$ is defined as a sensitivity coefficient. We need to clarify that the sensitivity to the true dipole, denoted as $n_{\sigma}=\mathcal{S_{\mathrm{true}}}\mathcal{A_{\mathrm{true}}}\sqrt{N}$, is not obtained directly from the reconstruction of the true dipole (or input dipole $\mathcal{A}$). Instead, it represents the corresponding true dipole of the $3\sigma$ or $5\sigma$ reconstructed dipole, denoted as $n_{\sigma}=\mathcal{S_{\mathrm{reco}}}\mathcal{A_{\mathrm{reco}}}\sqrt{N}$, which will be obtained from the observation by the IceCube-Gen2 surface array in the future. We scatter the data points $(n_{\sigma}/\sqrt{N}, \mathcal{A})$ for both the reconstructed dipole and true dipole cases and fit these data points with a polynomial equation of the form $\Sigma^3_{i=1}\lambda_{i}\mathcal{A}^{i}$, where $\lambda_{i}$ are the coefficients of the polynomial (see Figure \ref{fig4}). The slope of these curves represents the sensitivity coefficients for both the reconstructed and true dipole cases.
%%%%%% Equation 7 %%%%%%
\begin{equation}\label{eq7}
\begin{split}
\mathcal{S}(\mathcal{A}, E)=\lambda_{1}(E)+2\lambda_{2}(E)\mathcal{A}+3\lambda_{3}(E)\mathcal{A}^2.
\end{split}
\end{equation}
where $\lambda_{1}$, $\lambda_{2}$, $\lambda_{3}$, are different parameters for proton and iron which are energy dependent. Figure \ref{fig5} shows the ratio between the reconstructed dipole and the true dipole. 
%%%%%% Table2 %%%%%%
\begin{table}[b]
\centering
\footnotesize
\begin{tabular}{l c c c c c c c c} 
 \hline\hline
 Median energy $E_{i}$ & 1.8 PeV & 4.4 PeV & 11 PeV & 28 PeV & 70 PeV & 176 PeV & 441 PeV\\
 \hline
 $\mathcal{A_{\mathrm{reco}}}$ $(\times10^{-3})$ $3\sigma$ & 0.0865 & 0.1991 & 0.4934 & 1.2723 & 3.2468 & 8.5073 & 22.9160\\ 
              {} & 0.4998 & 1.1515 & 2.8600 & 7.4196 & 19.2255 & 52.6080 & 163.7749\\ 
  $\mathcal{A_{\mathrm{reco}}}$ $(\times10^{-3})$ $5\sigma$  & 0.1441 & 0.3318 & 0.8222 & 2.1203 & 5.4098 & 14.1687 & 38.0990\\ 
              {} & 0.8334 & 1.9210 & 4.7783 & 12.4455 & 32.5946 & 92.2597 & 341.6183\\
  $\mathcal{A_{\mathrm{true}}}$ $(\times10^{-3})$ $3\sigma$  & 0.5442 & 1.2528 & 3.1048 & 8.0071 & 20.4322 & 53.5486 & 144.4248\\ 
  $\mathcal{A_{\mathrm{true}}}$ $(\times10^{-3})$ $5\sigma$  & 0.9071 & 2.0880 & 5.1744 & 13.3435 & 34.0472 & 89.2095 & 241.9074\\ 

 \hline
\end{tabular}
\caption{Points of the $3\sigma$ and $5\sigma$ sensitivity bands for the reconstructed dipole for each energy bin $E_{i}$ ($i=$ 1 to 7), upper and lower boundary of the corresponding true dipole for both cases (sensitivity bands).}
\label{tab2}
\end{table}
The sensitivity curves and bands for both $3\sigma$ and $5\sigma$ are shown in Figure \ref{fig6}, the corresponding points are in Table~\ref{tab2}.

The sensitivity curves of the reconstructed dipole and bands of the corresponding true dipole (full dipole) of the IceCube-Gen2 surface array are shown in Figure \ref{fig6}, which represents the significance of a deviation from the null hypothesis (isotropy). The curves of $\mathcal{A_{\mathrm{reco}}}$ are lower than the bands of true amplitude due to the field of view (FoV) of the IceCube-Gen2 surface array. Taking into account the energy gap region with upper limits and the highest energy data from IceCube, the energy range chosen for this work is from $10^6$ GeV to the overlap region with Auger, approximately around $10^{8.8}$ GeV. A dipole with a very inclined orientation is difficult to reconstruct using the 1D projection method. Thus, IceCube-Gen2 will need to be combined with experiments on the norther hemisphere (e.g. IceCube \cite{Aartsen:2016au}, HAWC \cite{Abeysekara:2018uu}). Moreover, it is necessary to consider very inclined dipole declinations, due to the lack of information on dipole declination from existing data.

%%%%%% Figure 5 & 6 %%%%%%

%%%%%%%%%%%%%%%%% figure 7 %%%%%%%%%%%%%%%%%%%%
\begin{figure}[tbp]
\centering
\includegraphics[width=0.66\linewidth]{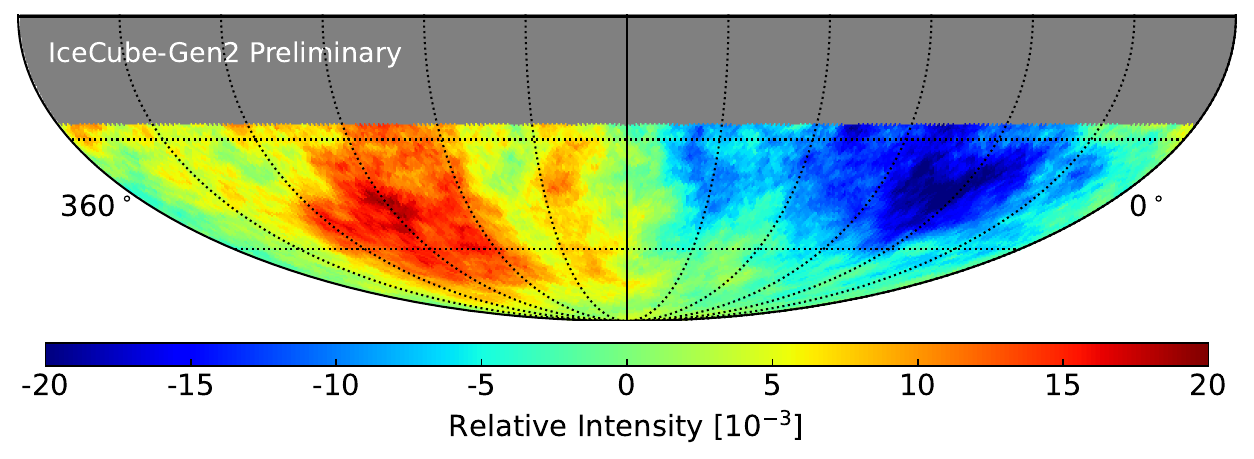}
\caption{Simulated relative intensity map for IceCube-Gen2 surface array with 5$\sigma$ of a reconstructed amplitude $9.029\times10^{-3}$ at 176 PeV. The injected dipole orientation is set as $(270\degree, -10\degree)$ for the right ascension and declination. The corresponding amplitude of true dipole is $1.437\times10^{-2}$.}
\label{fig7}
\end{figure}

By considering Eq. (\ref{eq6}) and (\ref{eq7}), we also plot the sky map of relative intensity at 176 PeV with an assumed $5\sigma$ significance level for the reconstructed dipole with amplitude of $9.029\times10^{-3}$, and with $20\degree$ top hat smoothing. The dipole orientation is set as $(270\degree, -10\degree)$ for the right ascension and declination. The corresponding amplitude of the injected true dipole ($\mathcal{A}=1.437\times10^{-2}$) can be derived using the reconstruction ratio $\mathcal{A_\mathrm{reco}}/\mathcal{A}$. We use HEALPix (with Nside = 64) for the sky map, the size of each pixel tile in the sky map is approximately $(0\degree.84)^2$.

%%%%%%%%%%%%%%%%%%%%%%%%%%%%%%%%%%%%%%%% Section 5 %%%%%%%%%%%%%%%%%%%%%%%%%%%%%%%%%%%%%%%%%%
\section{Conclusion and outlook}\label{sec5}

We present the 2D function of the air shower reconstruction efficiency of proton and iron primaries for the IceCube-Gen2 surface array based on the existing CORSIKA simulation, estimate the reconstruction efficiencies for helium, nitrogen, and aluminum using the natural logarithm of their mass numbers and show the total efficiencies of all particles utilizing the H4a flux model. Next, we simulate the CR arrival directions for the surface array using Monte-Carlo simulation by injecting 15 different dipoles at declinations ranging from $-80\degree$ to $80\degree$ for 7 energy bins, which range from $10^6$ GeV to $10^{8.8}$ GeV with a bin size of 0.4 in $\log_{10}(E/\mathrm{GeV})$. In total, we have 1785 injected dipoles. In order to study the anisotropy, we compare the actual sky map of an injected dipole with a reference map without a dipole for all 1785 cases, make the sky maps of relative intensity and perform 1D projections of the sky maps and fit them with first harmonic functions. To assess the sensitivity of IceCube-Gen2 surface array to a dipole anisotropy, we consider a null hypothesis and the propagation of the sigmas. Finally, we get the sensitivity function and show the $3\sigma$ and $5\sigma$ sensitivity to the CR dipole anisotropy with curves and bands for the surface array.

In order to improve the accuracy of the sensitivity analysis for the IceCube-Gen2 surface array, we will extend the threshold of the zenith angle in the CORSIKA simulation up to 80 degrees, which can provide a more accurate estimation of the reconstruction efficiency.

% Bibtex references:
\bibliographystyle{ICRC}
\bibliography{references}

% Alternatively, you can include references by hand:
%\begin{thebibliography}{99}
%\bibitem{...}
%
%\end{thebibliography}

\clearpage

%The following list of authors, affiliations and funding agencies will be updated at the day of submission. The following template is a placeholder generated via https://authorlist.icecube.wisc.edu/icecube on March 25, 2023 and will be updated.
\section*{Full Author List: IceCube-Gen2 Collaboration}

\scriptsize
\noindent
R. Abbasi$^{17}$,
M. Ackermann$^{76}$,
J. Adams$^{22}$,
S. K. Agarwalla$^{47,\: 77}$,
J. A. Aguilar$^{12}$,
M. Ahlers$^{26}$,
J.M. Alameddine$^{27}$,
N. M. Amin$^{53}$,
K. Andeen$^{50}$,
G. Anton$^{30}$
% ...
\\
\\
% ...
$^{12}$ Universit{\'e} Libre de Bruxelles, Science Faculty CP230, B-1050 Brussels, Belgium \\
% ...
$^{17}$ Department of Physics, Loyola University Chicago, Chicago, IL 60660, USA \\
% ...
$^{22}$ Dept. of Physics and Astronomy, University of Canterbury, Private Bag 4800, Christchurch, New Zealand \\
% ...
$^{26}$ Niels Bohr Institute, University of Copenhagen, DK-2100 Copenhagen, Denmark \\
$^{27}$ Dept. of Physics, TU Dortmund University, D-44221 Dortmund, Germany \\
% ...
$^{30}$ Erlangen Centre for Astroparticle Physics, Friedrich-Alexander-Universit{\"a}t Erlangen-N{\"u}rnberg, D-91058 Erlangen, Germany \\
% ...
$^{47}$ Dept. of Physics and Wisconsin IceCube Particle Astrophysics Center, University of Wisconsin{\textendash}Madison, Madison, WI 53706, USA \\
% ...
$^{50}$ Department of Physics, Marquette University, Milwaukee, WI, 53201, USA \\
% ...
$^{53}$ Bartol Research Institute and Dept. of Physics and Astronomy, University of Delaware, Newark, DE 19716, USA \\
% ...
$^{76}$ Deutsches Elektronen-Synchrotron DESY, Platanenallee 6, 15738 Zeuthen, Germany  \\
$^{77}$ Institute of Physics, Sachivalaya Marg, Sainik School Post, Bhubaneswar 751005, India \\
% ...

\subsection*{Acknowledgements}

\noindent
The authors gratefully acknowledge the support from the following agencies and institutions:
USA {\textendash} U.S. National Science Foundation-Office of Polar Programs,
U.S. National Science Foundation-Physics Division,
U.S. National Science Foundation-EPSCoR,
Wisconsin Alumni Research Foundation,
Center for High Throughput Computing (CHTC) at the University of Wisconsin{\textendash}Madison,
Open Science Grid (OSG),
Advanced Cyberinfrastructure Coordination Ecosystem: Services {\&} Support (ACCESS),
Frontera computing project at the Texas Advanced Computing Center,
U.S. Department of Energy-National Energy Research Scientific Computing Center,
Particle astrophysics research computing center at the University of Maryland,
Institute for Cyber-Enabled Research at Michigan State University,
and Astroparticle physics computational facility at Marquette University;
Belgium {\textendash} Funds for Scientific Research (FRS-FNRS and FWO),
FWO Odysseus and Big Science programmes,
and Belgian Federal Science Policy Office (Belspo);
Germany {\textendash} Bundesministerium f{\"u}r Bildung und Forschung (BMBF),
Deutsche Forschungsgemeinschaft (DFG),
Helmholtz Alliance for Astroparticle Physics (HAP),
Initiative and Networking Fund of the Helmholtz Association,
Deutsches Elektronen Synchrotron (DESY),
and High Performance Computing cluster of the RWTH Aachen;
Sweden {\textendash} Swedish Research Council,
Swedish Polar Research Secretariat,
Swedish National Infrastructure for Computing (SNIC),
and Knut and Alice Wallenberg Foundation;
European Union {\textendash} EGI Advanced Computing for research;
Australia {\textendash} Australian Research Council;
Canada {\textendash} Natural Sciences and Engineering Research Council of Canada,
Calcul Qu{\'e}bec, Compute Ontario, Canada Foundation for Innovation, WestGrid, and Compute Canada;
Denmark {\textendash} Villum Fonden, Carlsberg Foundation, and European Commission;
New Zealand {\textendash} Marsden Fund;
Japan {\textendash} Japan Society for Promotion of Science (JSPS)
and Institute for Global Prominent Research (IGPR) of Chiba University;
Korea {\textendash} National Research Foundation of Korea (NRF);
Switzerland {\textendash} Swiss National Science Foundation (SNSF);
United Kingdom {\textendash} Department of Physics, University of Oxford.

\end{document}